\documentclass[12pt,preprint]{aastex}
\usepackage{amsmath}
\usepackage{color}
\usepackage{rotating}

\shorttitle{Morphological evolution of 3D CME cloud }
\shortauthors{Feng et al.}

\begin{document}

\title{Morphological evolution of a 3D CME cloud reconstructed from three viewpoints}

\author{L. {Feng}\altaffilmark{1,2}, B. {Inhester}\altaffilmark{2}, 
        Y. {Wei}\altaffilmark{2}, W.Q. {Gan}\altaffilmark{1}, T.L. {Zhang}\altaffilmark{3},
        M.Y. {Wang}\altaffilmark{4}}
\email{lfeng@pmo.ac.cn}
\altaffiltext{1}{Key Laboratory of Dark Matter and Space Astronomy, 
  Purple Mountain Observatory, Chinese Academy of Sciences, 210008 Nanjing, China}
\altaffiltext{2}{Max-Planck-Institut f\"{u}r Sonnensystemforschung, Max-Planck-Str.2,
37191 Katlenburg-Lindau, Germany}
\altaffiltext{3}{Space Research Institute, Austrian Academy of Sciences, 8042 Graz, Austria}
\altaffiltext{4}{Shanghai Astronomical Observatory, Chinese Academy of Sciences, 
200030 Shanghai, China}

\begin{abstract}
The propagation properties of coronal mass ejections (CMEs) are crucial to predict its
geomagnetic effect. A newly developed three dimensional (3D) mask fitting reconstruction 
method using coronagraph images from three viewpoints has been described and applied to 
the CME ejected on August 7, 2010. The CME's 3D 
localisation, real shape and morphological evolution are presented. Due to its interaction
with the ambient solar wind, the morphology of this CME changed significantly in the 
early phase of evolution. Two hours after its initiation, it was expanding almost 
self-similarly. CME's 3D localisation is quite helpful to link remote sensing observations to 
in situ measurements. The investigated CME was propagating to Venus with its flank
just touching STEREO B. Its corresponding ICME in the interplanetary 
space shows a possible signature of a magnetic cloud with a preceding 
shock in VEX observations, while from STEREO B only a shock is observed. We 
have calculated three principle axes for the reconstructed 3D CME
cloud. The orientation of the major axis is in general consistent with the orientation of 
a filament (polarity inversion line) observed by SDO/AIA and SDO/HMI. The flux rope axis derived
by the MVA analysis from VEX indicates a radial-directed axis orientation. It might be
that locally only the leg of the flux rope passed through VEX. The height and speed
profiles from the Sun to Venus are obtained. We find that the CME speed 
possibly had been adjusted to the speed of the ambient solar wind flow after leaving 
COR2 field of view and before arriving Venus. A southward
deflection of the CME from the source region is found from the trajectory of the CME
geometric center. We attribute it to the influence of the coronal hole 
where the fast solar wind emanated from.  

\end{abstract}


\section{Introduction}

Coronal mass ejections (CMEs) are huge explosions of magnetized plasma from the Sun.
Their interplanetary counterparts, interplanetary coronal mass ejections (ICMEs), are
considered to be the main driver of geomagnetic storms \citep{Gonzalez:etal:1994}.
The major scientific objective of the \textit{Solar TErrestrial RElations Observatory} 
(STEREO) mission \citep{Kaiser:etal:2008} launched in October 2006 is to better 
understand the initiation and propagation of CMEs. A big advantage of the STEREO twin 
spacecraft is that they allow a simultaneous observation of a CME from two different
perspectives. Quite a number of papers have been dedicated to the 3D reconstructions
of CMEs based on the coronagraph observations of the STEREO spacecraft.

Different reconstruction methods of coronagraph observations have
been used in the past. Among them are forward modeling, triangulation
method, polarisation ratio method. A review on these
methods can be found in, e.g., \citet{Mierla:etal:2010}.
\citet{Antunes:etal:2009} used a combination of forward and inverse
method to estimate the CME mass distribution.

In this paper, we are aiming to obtain the 3D morphology of a CME by using the
coronagraph observations from three viewpoints. Unlike forward modeling where
a restricted family of geometrical CME shapes is assumed beforehand, our
method allows any a-priori shape of a CME.
For tie-point approaches, the identification of corresponding structures from
different views is required but is often difficult to achieve. The method
described in the current work does not have this problem.
The polarisation ratio method utilises the polarising properties of the
Thomson scattering. In principle, only one view is required. 
However, the
information returned is only a depth surface representative of the
scattering centres in the CME cloud along the line-of-sight. No information
about the depth distribution can be retrieved. Moreover, the resulting plane
bears the ambiguity of two symmetric solutions with respect to the plane of
sky. The combination of forward and inverse methods of \citet{Antunes:etal:2009}
also uses a forward modelling step essentially based on family of flux rope CME
models. In a second step, the flux rope 3D volume is then used as the
support for an inverse estimation of the density distribution of the CME.
The latter step, however, is heavily under constrained and multiple solutions
are possible.

Along with the very large variety of coronagraph reconstruction
techniques, 3D geometric reconstructions using heliospheric imagers have also
been developed by some authors, where a CME was treated as a complete volume 
rather than a point source. For example, \citet{Lugaz:etal:2009} used a 3D 
magneto-hydrodynamic code to disentangle observational from physical effects.
By comparing the simulation with the observations from coronagraph and 
heliospheric images,
the 3D nature of two successive CMEs and their evolution in the inner heliosphere
were studied. \citet{Tappin:Howard:2009} compared their model of interplanetary
disturbances with heliospheric image data from the Solar Mass Ejection Imager.
By identifying the simulated ICME that best matched the observations, they
obtained the parameters that can describe ICME's 3D leading-edge structure,
orientation and kinematics.

The CME localisation and morphology in 3D yield information about the CME's
orientation, propagation direction, angular width in longitude and latitude
etc. and help to make reliable predictions about the arrival time 
of a CME at Earth or other planets. Reversely, in-situ observations can
be combined with near-Sun 3D reconstructions to constrain the CME model
a-posteriori.
Such investigations aimed to study CME propagation properties in the
interplanetary space have been performed by some authors, e.g.,
\citet{Mostl:etal:2009, Byrne:etal:2010, Liu:etal:2010a, Liu:etal:2010b,
Rodriguez:etal:2011}. \citet{Rouillard:etal:2009} investigated a CME observed
between Sun and Venus using the STEREO, \textit{Venus Express} (VEX) and
MESSENGER data. They compared the (I)CME orientation obtained by white light
analysis with the in situ flux rope axis by MESSENGER and VEX. The CME
orientation in the STEREO/COR2 field of view (FOV) was derived by a fit of the
graduated cylindrical shell (GCS) model \citep{Thernisien:etal:2006,
Thernisien:etal:2009, Thernisien:2011}. Due to the limited separation angle at
the time of these observations, the GCS flux rope was only fitted to COR2 A.
Early work before the STEREO era devoted to the comparison of the orientation
of the source region neutral line, the CME cloud shape in coronagraph images
and the flux rope axis orientation from in situ measurements have been made by
\citet{Yurchyshyn:etal:2001, Cremades:Bothmer:2004, Yurchyshyn:2008}.

In this paper, we propose a new method for the reconstruction of the
CME cloud based on the back projection of the observed CME periphery in
multiple coronagraph images without any restriction by a predefined
class of CME shapes.
From the superposition of these back projections, we obtain a 3D volume which
must contain the CME cloud entirely. Reversely, the real CME cloud also must
fill this 3D volume in the sense that the cloud has to touch each wall
of the 3D volume somewhere. Based on these principles, we propose a scheme
to obtain an approximation to the real 3D CME cloud shape.
Since the problem of 3D reconstruction from 2D images is notoriously
under constrained, we have designed our scheme so as to incorporate
images from as many different view directions as possible. In the
application presented here, coronagraph data from three spacecraft
were used.

The paper is organised as follows:
In \S 2, the coronagraph observations from three vantage points, STEREO A \& B,
and SOHO are described.
Details of our reconstruction method are presented in \S 3.
In \S 4, we use the 3D morphology obtained from our reconstruction to
interpret in-situ magnetic field and plasma parameters observed during the
passage of the CME near Venus, Earth, and two STEREO spacecraft. The data we 
consider were obtained by
the magnetometers (MAG) onboard VEX orbiting Venus, \textit{Advanced Composition 
Explorer} (ACE) orbiting Earth, and from
the In-situ Measurements of Particles and CME Transients (IMPACT) and The Plasma and 
Suprathermal Ion Composition (PLASTIC) telescopes onboard STEREO.
From a series of 3D reconstructions with time, we are able to follow the CME
morphological evolution. The height, speed evolution from the Sun to Venus
are also discussed.
The orientation of the filament in the source region, the orientation of the
CME major principle axes and the flux rope axis direction derived from VEX/MAG
data are compared.

\section{Observations and data reduction}

\begin{figure} 
  \centering
  \includegraphics[width=15.cm, height=12.5cm]{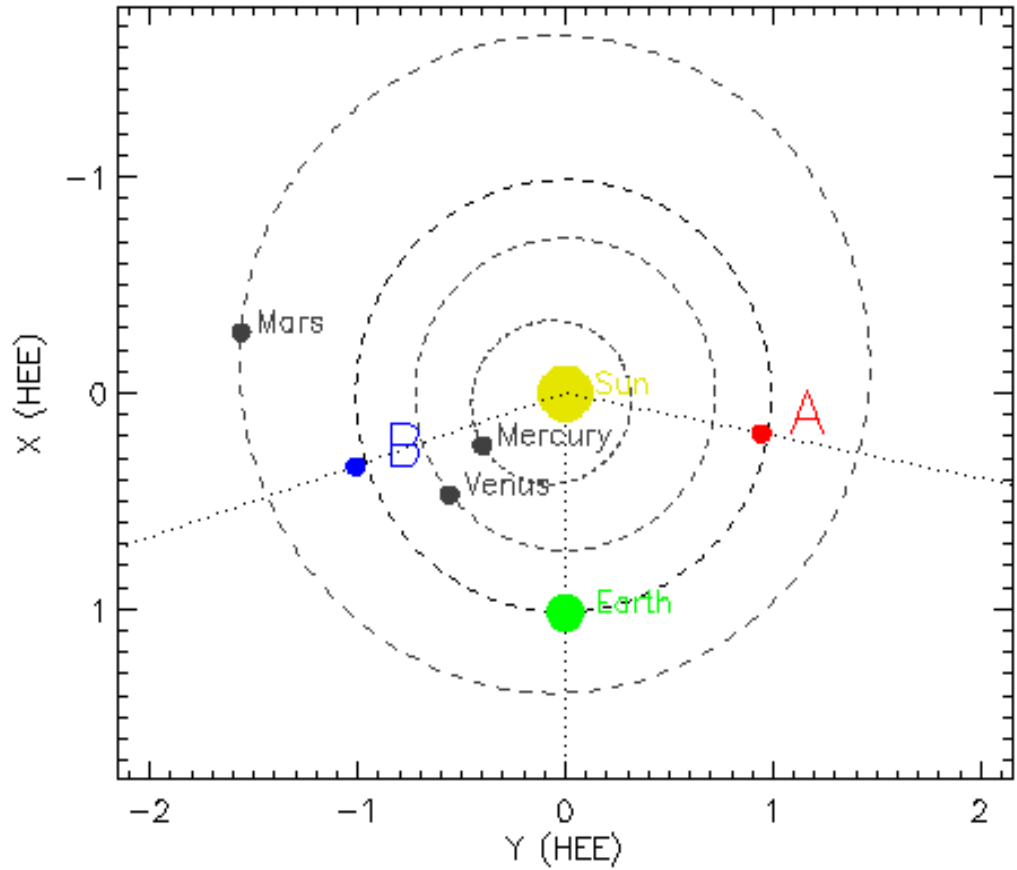}
  \caption{Relative positions of two STEREO spacecraft, planets and their 
     orbits in the inner solar system in the frame of HEE coordinate system.}
  \label{fig:HEE_posi}
\end{figure}

The CME we have investigated was observed on August 7, 2010 from three 
perspectives by the white light coronagraphs 
Large Angle Spectroscopic Coronagraph
(LASCO; \citet{Brueckner:etal:1995}) C2 and C3 onboard the \textit{Solar and 
Heliospheric Observatory} (SOHO; \citet{Domingo:etal:1995}), COR1 and COR2 in the 
the Sun Earth Connection Coronal and Heliospheric
Investigation (SECCHI; \citet{Howard:etal:2008}) instrument suite onboard STEREO.
LASCO C2 has a FOV from 2 to 6~R$_{\odot}$, C3 from 3.7 to 
30~R$_{\odot}$. COR1 reaches a lower altitude with its FOV from 1.4 to 
4~R$_{\odot}$, COR2 from 2.5 to 15~R$_{\odot}$. 
On August 7, two STEREO spacecraft were separated by around 150 degrees.
The spatial positions of STEREO A and B, the planets and their orbits in the inner
solar system are presented in Figure~\ref{fig:HEE_posi} in the Heliocentric Earth
Ecliptic (HEE) coordinate system. Venus was located between STEREO B and the
Earth. The magnetometer onboard VEX \citep{Zhang:etal:2006} will be utilised to verify the CME 
propagation direction and to determine the CME arrival time at Venus. 

\begin{sidewaysfigure} 
  \centering                 
  \includegraphics[width=22cm, height=12.5cm]{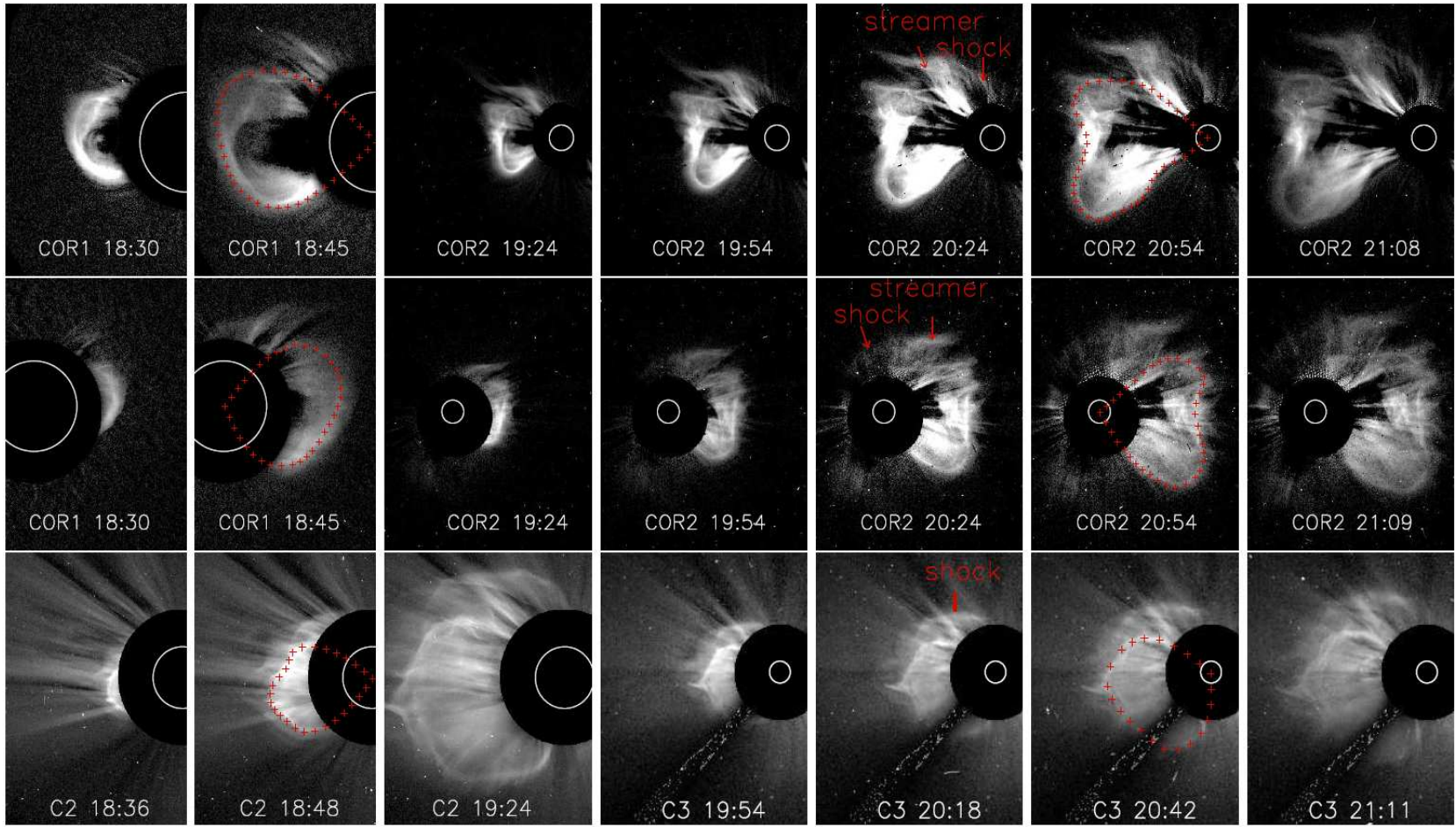}                  
  \caption{Selected coronagraph observations from three viewpoints. 
  The top row is COR1 and COR2 images from STEREO A, the middle row
  is the COR1 and COR2 images from STEREO B and the bottom row is the
  LASCO C2 and C3 images from SOHO. The white circle in each panel indicates
  the limb of the solar disk and the dark round area is caused by the coronagraph 
  occulter. For COR data, the respective pre-cme images are subtracted; 
  while for LASCO data, a 12-hour minimum image is created and subtracted 
  for C2 and C3. The small red plus signs mark the positions of CME peripheries.}
  \label{fig:corAB_lasco}
\end{sidewaysfigure}

In Figure~\ref{fig:corAB_lasco} we present the time series of coronagraph
images of the CME event for the three viewpoints employed. All images were
processed by standard Solarsoft routines secchi\_prep and lasco\_prep.
COR1 and COR2 images from STEREO A are shown in the first row, COR 
images from STEREO B are in the second row, and the C2 and C3 images
from LASCO are presented in the last row.
In each panel, the operated instrument and observational time are marked at
the bottom. The occulter of coronagraph is indicated as a black mask. As a size
reference, we also plot the solar limb as a white circle onto each panel.
Two different background subtraction methods have been applied here to
make the CME leading edge and its related shock more prominent.
For COR data, the respective pre-cme images were subtracted. For LASCO-C2, a
12-hour minimum image was created by taking the minimum brightness of each
pixel during the 12 hours centred around the CME's first appearance.
The dark area inside the CME in COR1 and COR2 indicates the position of a
preexisting helmet streamer which stayed more or less quasi-stationary during
the CME evolution.
The first appearance of the CME in COR1 A was at 18:15, in
COR1 B at 18:25 UT, and in C2 at 18:36 UT.

For this particular CME, we identified shock signatures in the white-light 
observations. They appeared as a smooth front which outlined 
the CME out-most envelope and was 
associated, spatially and temporally, with streamer deflections 
\citep{Ontiveros:Vourlidas:2009}. They are marked in column 5 in 
Figure~\ref{fig:corAB_lasco} with two deflected streamers accompanied closely.
At lower altitudes as in column 2, the shock looked like a diffusive area 
enclosing the CME cloud \citep{Gopalswamy:Yashiro:2011,Gopalswamy:etal:2012}. It is best seen 
in the C2 image of column 2 where the CME periphery is
indicated by plus signs. The presence of a shock for this CME event is
confirmed by the type II radio burst at a height of about 1.36~R$_{\odot}$ (N.
Gopalswamy 2011, private communication). 

\begin{figure} 
\centering
 \includegraphics[width=16.5cm, height=11.cm]{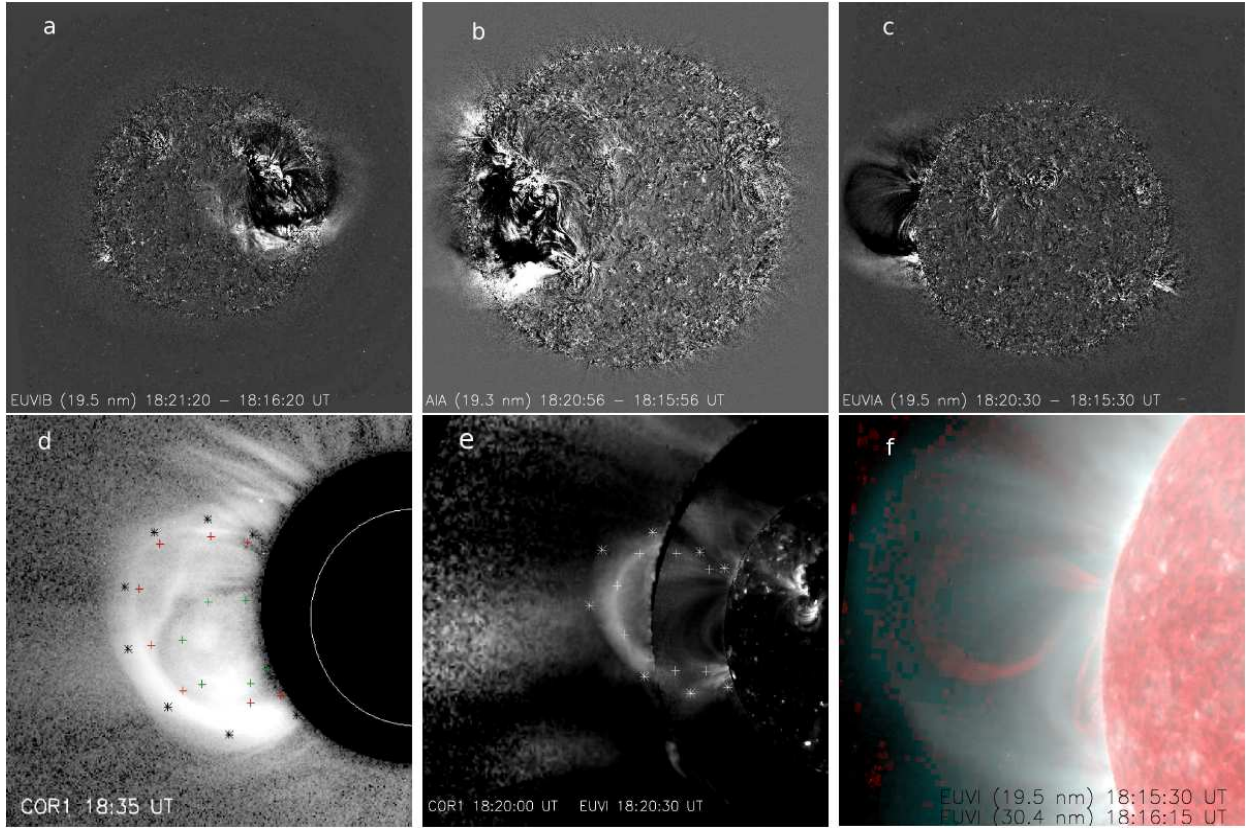}
 \caption{Upper: panels (a), (b), and (c) are the running difference
 images of the dimming region and the coronal EIT waves observed from EUVI B, AIA and EUVI A.   
 In panel (f) the red color channel represents the erupted prominence at 18:16 UT in 30.4 nm and 
 the light blue channel represents the plasma emission at 18:15 UT in 19.5 nm, where we
 can see the dome-shape wave overlying the prominence.
 In panel (e) the CME periphery and its preceding shock front are indicated in plus signs and 
 asterisk signs, respectively. In panel (d), the internal structure of the CME is presented.
  The black asterisk, red and green plus signs indicate the CME leading edge, cavity 
  outer boundary and core region, respectively. (An animation of the EIT wave in one-minute
  cadence AIA images is available in the online journal.)}
 \label{fig:cme_ini} 
\end{figure}

The shock signature can presumably be linked to the coronal EIT waves 
down to the FOV of EUVI \citep{Wuelser:etal:2004} onboard STEREO \ and AIA 
\citep{Lemen:etal:2011} onboard SDO (\textit{Solar Dynamics Observatory}),
which are 1.7 and 1.3 solar radii, respectively. In panel (a), (b), and (c) of
Figure~\ref{fig:cme_ini}, we present the dimming and the wave front
in the running difference images observed by EUVI B, AIA, and EUVI A. An
animation of the high-cadence AIA observations is available in the online 
journal where the coronal wave signature can be seen more prominently. The wave
signatures observed above the limb were well connected to the wave signatures
observed against the solar disk. EUVI A
provided a better view of the wave dome. Inside the dome, the erupted
prominence was visible in panel (f). The prominence in 30.4 nm is colored
in red, while the wave front is colored in light blue.  When the CME
entered the COR 1 FOV (panel (e)), the CME periphery and shock
front are delineated by plus signs and asterisk signs,
respectively. In the same panel, the CME associated streamer belt 
is also shown. Details of the shock/EIT wave analysis is
beyond the scope of this paper and will be pursued in a 
separate paper. Similar wave behavior and related quantative
analyses can be found in, e.g., \citet{Veronig:etal:2010}.

The subtraction of the pre-CME image somehow removed part of 
the internal structure of the CME. To make the internal structures visible, 
for COR 1 images we applied the 12-hour minimum image as well. The resulting 
images are shown in panel (d) of Figure~\ref{fig:cme_ini}. The CME periphery
we used for 3D reconstruction is 
indicated as black asterisk signs. We tried to exclude as well the sheath between 
the shock front and the CME which manifested as a diffusive region there. 
Red plus signs represent the outer boundary of the CME cavity and the area inside
green signs indicates the core region. For this CME, we found that
its internal structure is a little bit complicated. In the cavity, it seems that
a bright line feature appeared. 
If we regard the CME flux rope as the cavity and the prominence attached below 
\citep{Chen:1996}, the area inside the red signs is mainly the CME flux rope 
and possibly some material following it. Except the unknown part hidden behind the
occulter, we found that the CME flux rope occupied most of the area surrounded 
by the black signs, which we used for the 3D reconstructions. 

In order to locate the CME source region, we have checked EUV images at
various wavelengths taken by the EUVI instrument 
onboard STEREO and by the AIA telescope onboard the
SDO spacecraft. At the time of the CME
launch, a M1.0 class flare occurred in active region AR~11093 located at
N12E31 as viewed from Earth on August 7. According to GOES light curves, the
flare started around 17:55 and peaked at 18:24. Magnetograms from HMI
(Helioseismic and Magnetic Imager, \citet{Scherrer:etal:2011}) were used to
determine the orientation of the magnetic polarity inversion line in the CME
source region.

\section{3D reconstruction method}

\subsection{3D reconstruction and smoothing}

In the traditional stereoscopic technique, two corresponding points
from an image pair are back projected along the respective line-of-sight to
obtain their 3D coordinates
\citep{Inhester:2006,Feng:etal:2007a,Feng:etal:2009, Liewer:etal:2009,
Liewer:etal:2011}. \citet{deKoning:etal:2009} geometrically reconstructed
the CME location plane by plane from two viewpoints. The authors selected
two points on the CME boundary in each image and back-projected to the Sun
totally four points which lie in the same plane. Their back-projections
formed a quadrilateral in that specified plane where the CME is ideally
localised. Later, \citet{deKoning:Pizzo:2011} included the geometric localisation
method into the polarimetric technique to remove the ambiguity inherent
in this technique.  

The idea of our reconstruction method employs the inverse 
approach of the traditional stereoscopic technique. In this newly developed 
method, we forward project a point in the 3D space onto each coronagraph
image. If the three projections are all located inside the respective CME
periphery observed in each image, this 3D point is considered as a point
inside a 3D volume which must contain the 3D CME cloud entirely.
On the other hand, for the CME surface to be consistent with the observed
peripheries, it must be close enough to the boundary of the
reconstructed 3D volume so that their projections onto the coronagraph
images coincide.

In a first step, we identify the CME periphery in each coronagraph image.
At the moment, it is done by tie-pointing the CME leading edge in each 
image. The tie-points are then interpolated by a parametric cubic spline
to obtain a smooth periphery curve.
Examples of periphery tie-points are shown in column 2 and 
6 of Figure~\ref{fig:corAB_lasco}. Since we do not know how the CME
continues behind the occulter, we simply extrapolate the ends of the
periphery curve from the occulter edge radially to the Sun's surface. 
The result of this step is saved as binary masks $M_i(x_{p,i},y_{p,i})$
for each image $i=A,L,B$ taken by STEREO A, LASCO and STEREO B, respectively.
We set $M_i$ to unity at image pixels
$(x_{p,i},y_{p,i})$ inside the periphery curve and to zero outside.
Note that the LASCO images were not always recorded exactly simultaneously
to those from STEREO. In these cases, the CME periphery curves from
two neighbouring frames in time from LASCO were interpolated to the time
of the STEREO observations.

As the next step, we select a regular 3D mesh of the space around the Sun. In
our method, the distance of grid points is chosen adaptively depending on the
projected CME size in the images. Since we are considering a time evolution
problem, all calculations in this work are performed in a Cartesian Carrington
coordinate system.
For each 3D grid point $\mathbf{r}=[x,y,z]$ we obtain its projection
$\mathbf{r}_{p,i}$ onto image $i$ by \citet{Feng:etal:2007b}.
\begin{equation}
\mathbf{r}_{p,i}=A_i^T\mathbf{r}
\label{equ:project}
\end{equation}
where $A_i$ is the the matrix of the coordinate transform 
\[
A_i=
\left[ {\begin{array}{ccc}
 -\sin L_i & -\cos L_i\sin B_i \\
  \cos L_i & -\sin L_i\sin B_i \\
  0        &  \cos B_i         \\
 \end{array} } \right]
\]
and $L_i$ and $B_i$ are the longitude and latitude of spacecraft $i$ in the
Carrington coordinate system. After projecting a 3D point $\mathbf{r}$ onto
all three coronagraph masks $M_i$, we set a 3D mask $M_{3D}$ according to
\begin{equation}
M_{3D}(\mathbf{r})=\begin{cases}
   1 \quad\text{if}\;M_i(\mathbf{r}_{p,i})=1\;\text{for all}\;i=A,L,B\\
   0 \quad\text{if}\;M_i(\mathbf{r}_{p,i})=0\;\text{for one or more}\;i
\end{cases}
\end{equation}
This process is repeated for all 3D points until the 3D mask $M_{3D}$ is
completely determined. This mask then defines the embedding 3D volume
containing the CME cloud.

\begin{figure} 
\centering
\includegraphics[width=11.cm, height=10.cm]{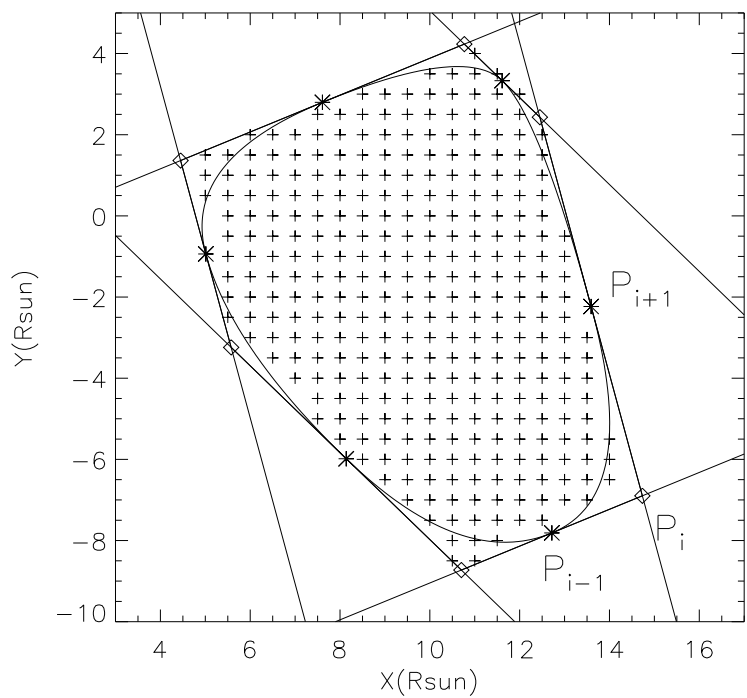}
\caption{Smoothing of the CME cloud by quadratic B\'{e}zier curves in one 
 horizontal plane parallel to the solar equatorial plane. The plus signs are 
 the 3D CME points lying in this plane from the mask fitting method.
 Three pairs of parallel lines are the projections
 of line of sights from STEREO A, B and SOHO onto this horizontal plane.
 Their intersections are shown by diamonds and the respective middle points are 
 marked by asterisks. The B\'{e}zier curves are shown by a smoothed consecutive
 curve which is tangent to the projected line of sights. }
\label{fig:bezier}
\end{figure}

As an example of the reconstructed 3D mask we show a horizontal cut of
$M_{3D}$ which yields the polygon in Figure~\ref{fig:bezier}. The mask area is
marked by plus signs at the respective grid points. The region defined by mask
$M_{3D}$ may include areas which do not belong to the true CME area but are
yet covered but the projections of the 2D image masks. Assuming the real CME
surface is smooth, these areas are probably confined to the vicinity of the
vertices of the polygon from $M_{3D}$. These vertices occur naturally where
the boundaries of the back projected 2D masks intersect. From three images
there are up to six vertices possible in every horizontal plane intersecting
$M_{3D}$. In Figure~\ref{fig:bezier}, these vertices are marked as diamonds.

In a final step, we smooth the corners of $M_{3D}$. For each horizontal plane,
we find the centres on the faces of the 3D mask between each neighbouring pair
of vertices. In Figure~\ref{fig:bezier}, these centres are marked as stars. For
each vertex we then define a quadratic B\'{e}zier spline
\begin{equation}
\mathbf{B_i}(t)=(1-t)^2\mathbf{P_{i-1}}+2(1-t)t\mathbf{P_i}+t^2\mathbf{P_{i+1}}
 ,\quad t\in[0,1]
\label{equ:bezier}
\end{equation}
from the three control nodes $P_{i-1}$, $P_i$ and $P_{i+1}$. The central
$P_i$ is the vertex position and $P_{i\pm1}$ are located at the face centres
to either side to the vertex. This choice of the spline curve insures that
the boundary formed by the combination of all spline curves is continuous
and smooth as demonstrated by the example in Figure~\ref{fig:bezier}.
This new boundary is finally adapted as the approximation to the CME shape
in each horizontal plane.

Similar techniques were used in earlier papers. \citet{Byrne:etal:2010} fitted
an ellipse to a quadrilateral which was derived from two viewpoints. However,
our B\'{e}zier curve fitting has more flexibility and is readily adapted to
include more than two view directions.
Additional view directions provide more constraints to the 3D CME localisation.
However, we have to admit that as the control nodes on the face centres are
selected somewhat arbitrarily, there is still some uncertainty about the
real CME shape. From our procedure we obtain the surface with the
least curvature compatible with the observations.

\begin{figure} 
\centering
\includegraphics[width=15.cm, height=12.cm]{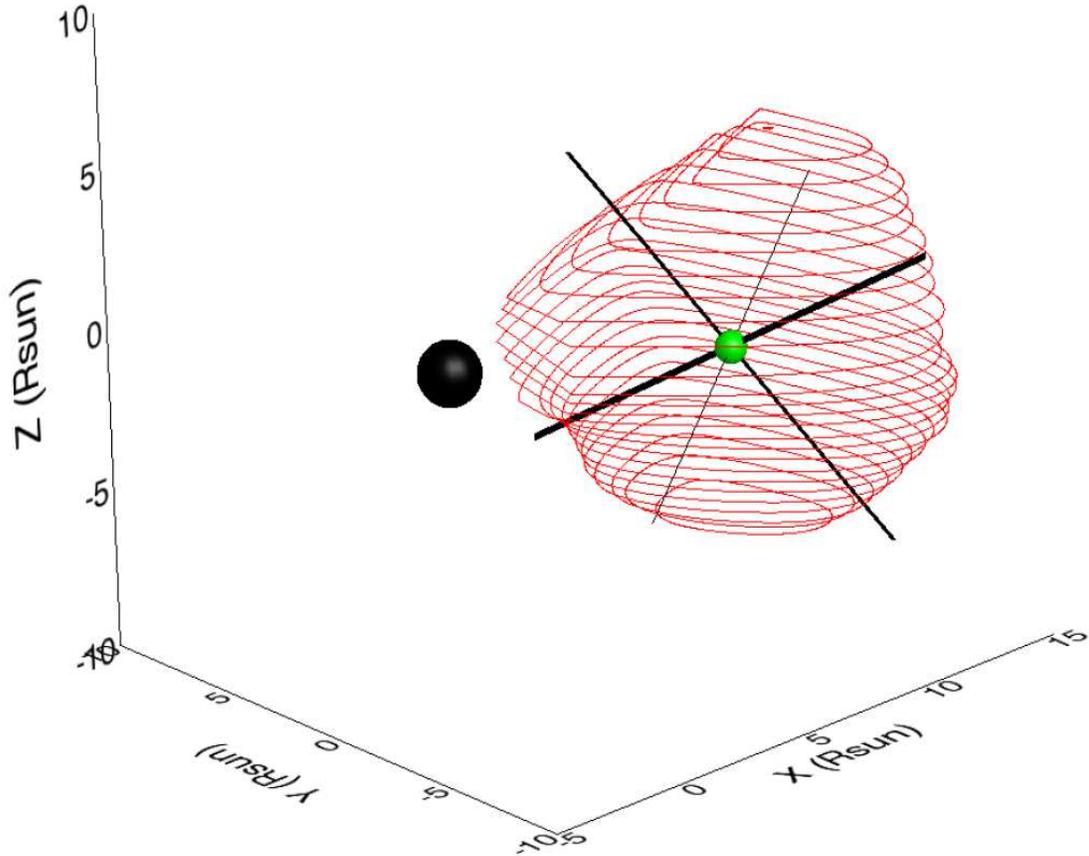}
\caption{One example of 3D CME cloud and its three principle axes at 21:24 UT.
The thickness of axis is proportional to the scale of eigenvalue. The
green dot indicates the position of CME's geometric centre. }
\label{fig:prin_axis_eg}
\end{figure}

This fitting process described above is performed for each horizontal plane.
In the end, a smoothed surface of the CME cloud is obtained. One example of
such reconstruction is shown in Figure~\ref{fig:prin_axis_eg} and is
represented by red curves. We have to mention that since the CME structure
below the occulter is unknown, the CME cloud in the region inside the
occulter radius from the Sun's centre is not plotted.

\subsection{Calculation of the geometric centre and principle axes}

Since a determination of the density distribution in the CME cloud
is a highly ill-posed problem and can not be done reliably from only
three view directions without further constraints, we take the CME
shape as the basis for a further analysis.
We therefore neglect the internal structure of the CME in the analysis
below and only characterize the shape of the cloud and its morphological
evolution. Hence the centre of the cloud may not be its centre of gravity, 
but the geometric centre which is still a significant
quantity to characterise the position and motion of the CME cloud.

Natural characteristics of an amorphous volume are its geometric centre
(GC), its three principle axes and their eigenvalues. The geometric centre
$\mathbf{r_{gc}}$ of the CME cloud can be obtained from
\begin{equation}
\mathbf{r_{gc}}=\frac{\sum_\mathbf{r} V(\mathbf{r})\mathbf{r}}
{\sum_\mathbf{r}V(\mathbf{r})}
\label{equ:cgrav} 
\end{equation}
where $V(\mathbf{r})$ is unity inside the CME cloud and zero elsewhere.
The principle axes and their eigenvalues are found from a diagonalisation
of matrix
\begin{equation} 
\frac{\sum_\mathbf{r}V(\mathbf{r})(\mathbf{r}-\mathbf{r_{gc}})
     (\mathbf{r}-\mathbf{r_{gc}})^{T}}
     {\sum_\mathbf{r}V(\mathbf{r})}
\end{equation}
These integrations are straight forwardly performed on the 3D grid chosen
to generate the mask $M_{3D}$. Grid cells intersected by the boundary of
$V(\mathbf{r})$ are weighted according to their overlap with $V(\mathbf{r})$.

In Figure~\ref{fig:prin_axis_eg}, we plot the directions of three eigenvectors
calculated from the matrix diagonalisation for the CME cloud at 21:24. The
thickness of three principle axes is proportional to the magnitude of
eigenvalues. All three eigenvectors are centred at CME's geometric centre.
We name minor, intermediate and major axis, respectively, according
to the order of their eigenvalues.
 
\section{Results}

Based on the reconstructed 3D surface of the CME cloud, further analyses can
be made. A straight forward application is to identify the longitude and
latitude of the CME cloud and its spatial extension in 3D. This information
can assist to interpret in-situ measurements. From a time series of the 3D
cloud, the CME's morphological evolution and propagation
properties can be derived. 

\subsection{Interpretation of the in situ data}

From the space weather point of view, the longitude and latitude
position, the angular width of the CME cloud are very crucial to determine
whether or not a CME can interact with a planet's magnetosphere or
a spacecraft.

\begin{figure} 
\centering
\includegraphics[width=12.cm, height=11.cm]{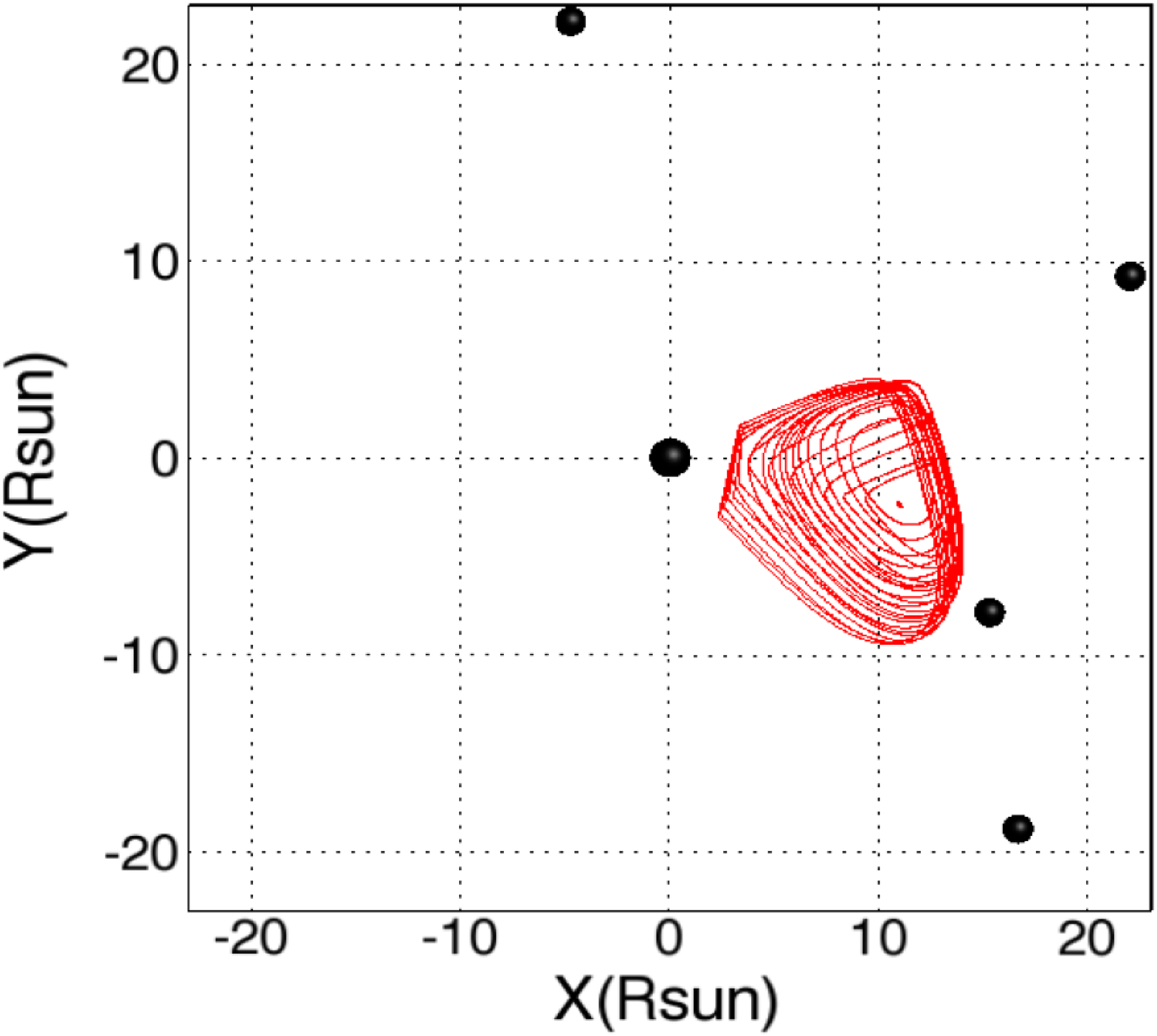}
\caption{A perspective from a vantage point above the north pole of the Sun to 
show the localisation
of the CME relative to the Sun, Venus, Earth and two STEREO spacecraft at 21:24 UT. The Sun 
centres at (0,0). From top to bottom, the other four black spheres represent STEREO 
A, Earth, Venus and STEREO B, respectively. Here their distances to the Sun have
been scaled to 10 \% of their original values to just incidate their view directions.}
\label{fig:topview}
\end{figure}

\begin{figure} 
 \centering
 \includegraphics[width=16.cm, height=0.5\textheight]{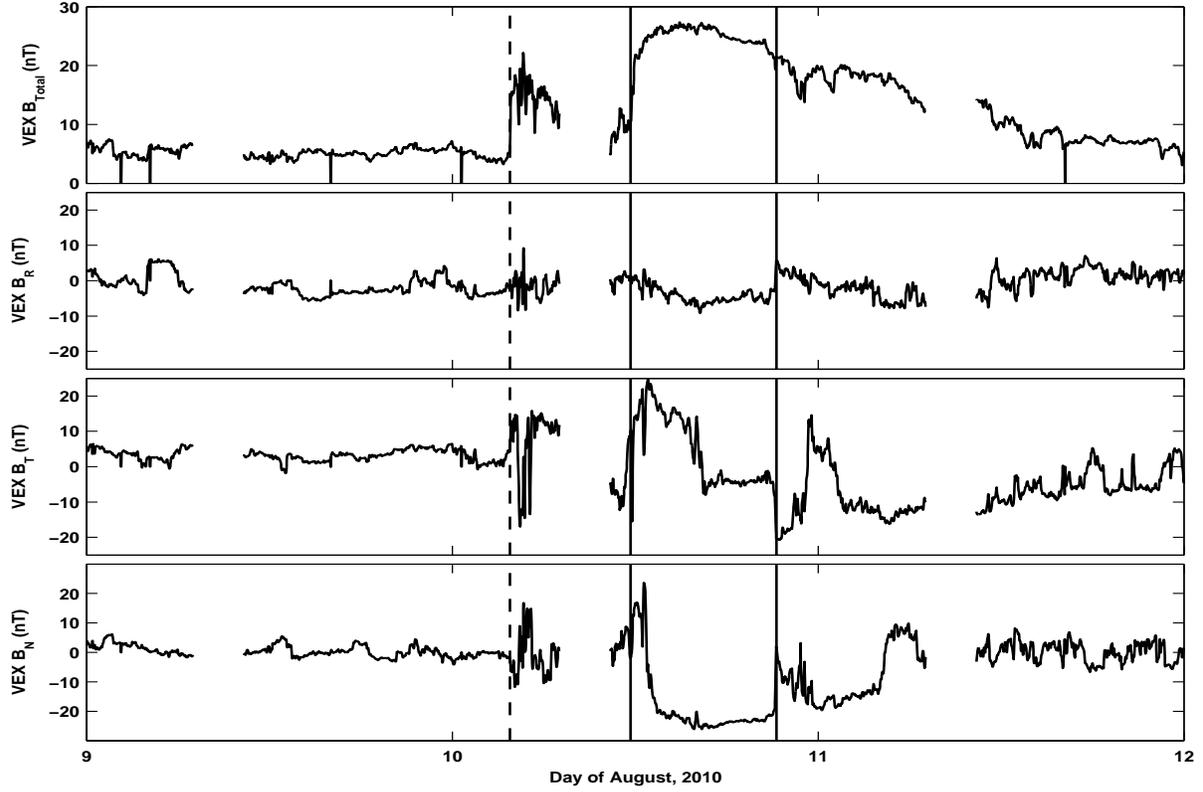}
 \caption{In situ measurements by MAG onboard Venus Express.
 From top to bottom are the total magnetic strength, three magnetic field components
 in RTN coordinates. The signature of magnetic cloud is indicated by two vertical
 solid lines and the preceding shock is marked by the dashed line.  VEX has an 
 elliptical polar orbit with a 24 hours period, and stays in Venusian magnetosheath 
 and magnetosphere from 07:00 to 10:20 UT for each orbit. 
 MAG continuously operates, but we block the data 
 from 07:00 to 10:20 UT to focus on the solar wind structure.}
 \label{fig:vex}
\end{figure}

Figure~\ref{fig:topview} provides a view from a vantage point above the 
solar north pole in which the localisation of CME relative to the Sun, 
Venus and two STEREO spacecraft are indicated. The CME initiated on August 7 was 
propagating closely towards Venus. In Figure~\ref{fig:vex} 
we present the magnetic field measurement by the Magnetometer onboard 
Venus Express (VEX/MAG) from August 9 to 12. From top to bottom, we 
plot the total magnetic strength
and three magnetic field components in the Radial-Tangential-Normal (RTN)
coordinate frame. Here, vector $R$ is the unit vector from Sun to the
spacecraft , $T$ is the unit vector in the direction of the cross product of
$R$ and solar rotational axis, and $N$ completes the right-handed triad. VEX
has an elliptical polar orbit with a 24-hour period, and stays in the Venusian
magnetosheath and magnetosphere from 07:00 to 10:20 UT for each orbit. MAG
continuously operates, but we block the data from 07:00 to 10:20 UT to focus
on the solar wind structure. 

On August 10, from about 04:00 UT MAG detected a sudden 
discontinuous increase in the magnetic field strength and the abrupt changes
in the field direction. It is likely the shock and sheath signature preceding 
an ICME. Further verification requires also a sudden increase in flow speed, 
density and thermal speed. Unfortunately, plasma measurements by VEX was
not available for the solar wind structure. The spacecraft enters the
Venusian ionosphere when plasma measurements are made. From 11:00 UT on the
same day, the magnetic field strength was enhanced and smooth rotations of 
the field were found. It implies the appearance of a magnetic cloud. Due to
the lack of plasma measurements, a further proof of the lower proton 
temperature is not available in Figure~\ref{fig:topview}.

\begin{figure} 
 \centering
 \includegraphics[width=12.cm, height=\textheight]{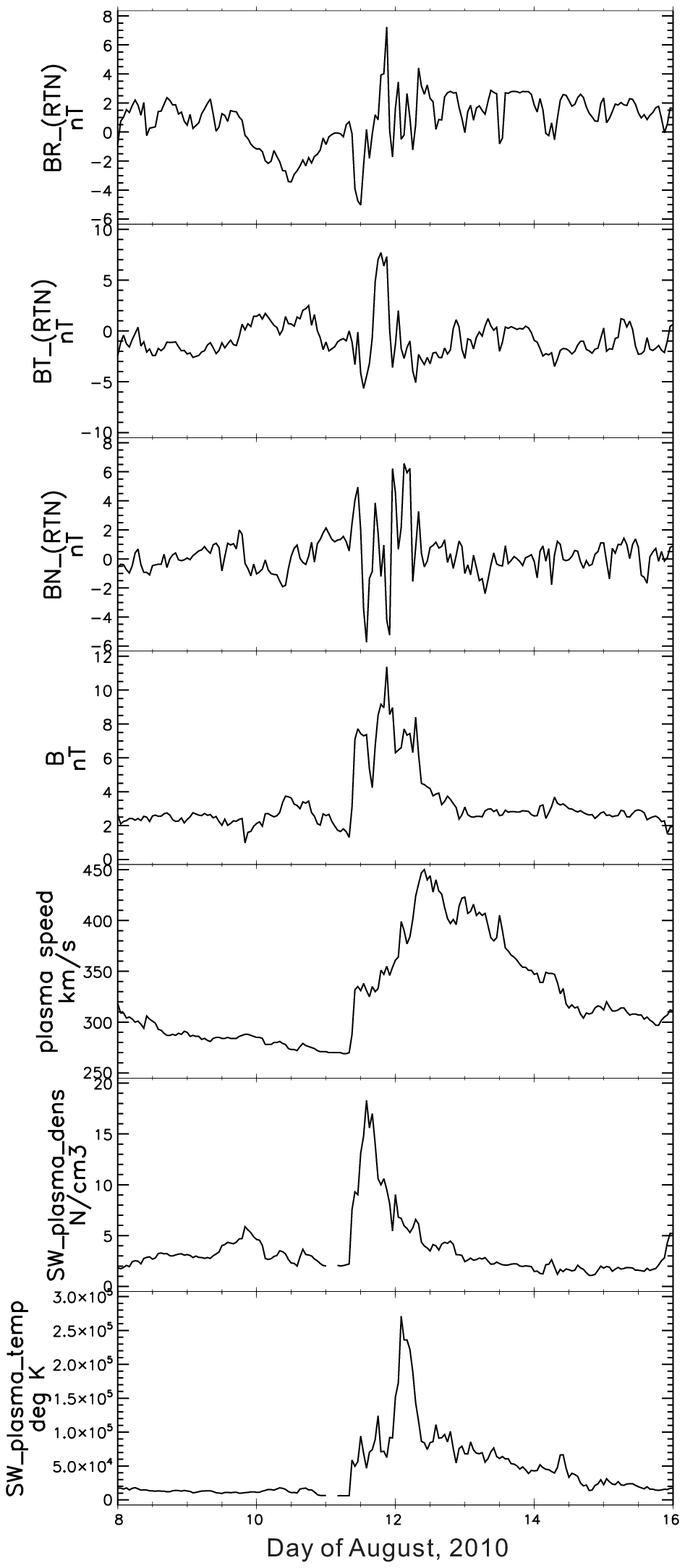}
 \caption{IMPACT and PlASTIC data with a temporal resolution of one hour 
 from STEREO B. From top to bottom are the 
 magnetic field components in RTN coordinates and its total strength, the proton density,
 bulk speed and proton temperature.}
 \label{fig:insituB}
\end{figure}

The magnetic field and plasma measurements in Figure~\ref{fig:insituB} taken 
from the IMPACT and PLASTIC particle spectrometers onboard STEREO B only show the
signature of a shock which appeared around 10:00 UT on August 11. From top to
bottom, we plot the magnetic field in RTN coordinates and its total strength,
the proton density, the bulk speed and the proton temperature.
We did not see any clear ICME at STEREO B following the shock. This observation
can be explained by checking the CME localisation relative to the line
connecting the Sun and STEREO B in Figure~\ref{fig:topview}.
The flank of the reconstructed CME cloud just touches this connection line
which is an evidence that the cloud did not pass STEREO B. However, since the shock
has a larger spatial extension than the cloud, it was still observed at the
spacecraft. The projected shape of the CME on the solar 
equatorial plane indicated that the CME was mainly directed to VEX and STEREO B.
Unfortunately, we do not have direct observations of the shock on the
equatorial plane, we speculate that the in-situ shock signature might be more prominent 
in the VEX and STEREO B observations. We did not see any signature of an ICME 
and/or a shock in the ACE data observed at L1 point.

\subsection{Morphological evolution of the 3D cloud}

\begin{figure} 
  \centering
  \vbox{
  \includegraphics[width=14.cm, height=7cm]{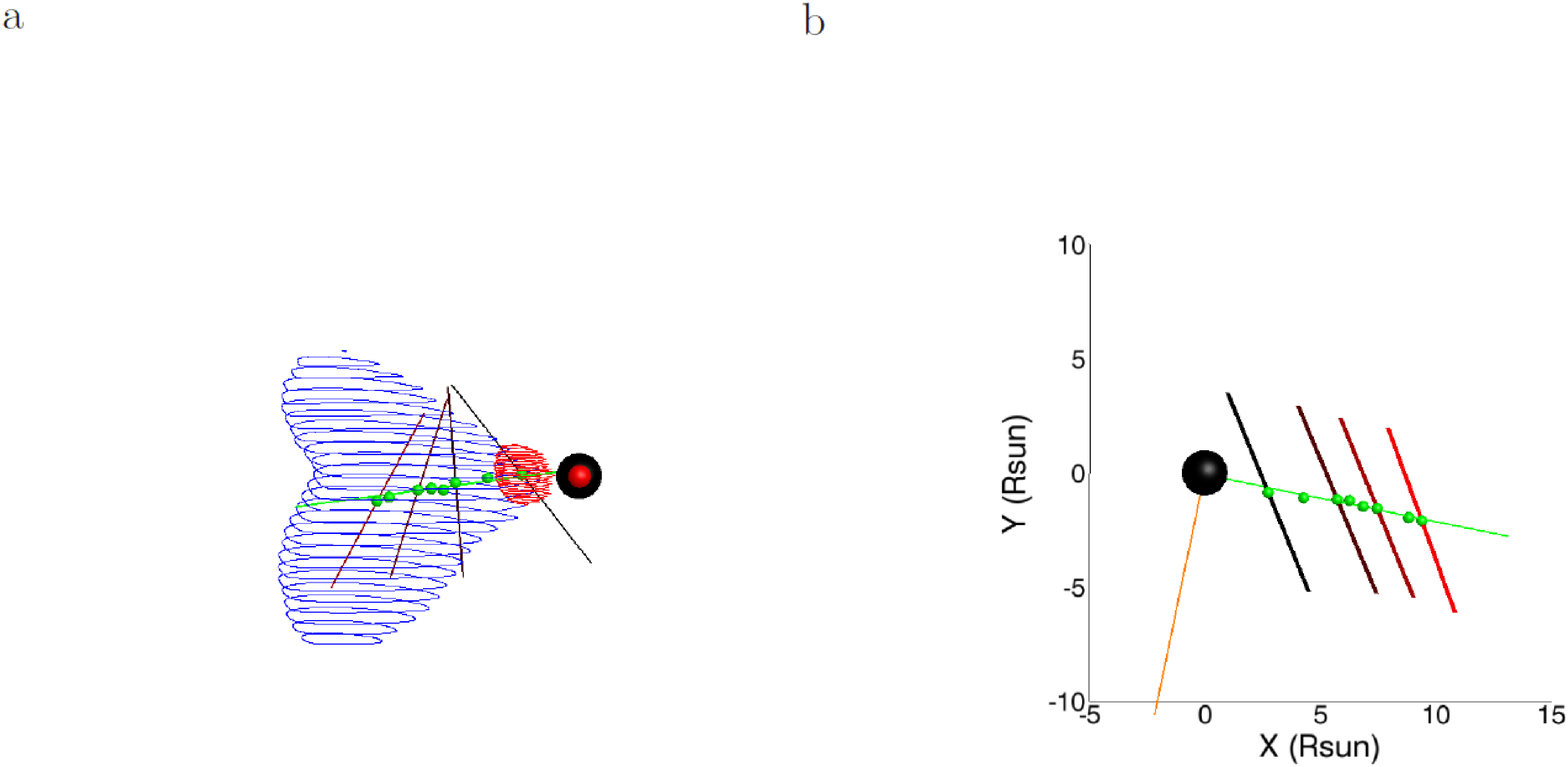}
  \includegraphics[width=14.cm, height=7cm]{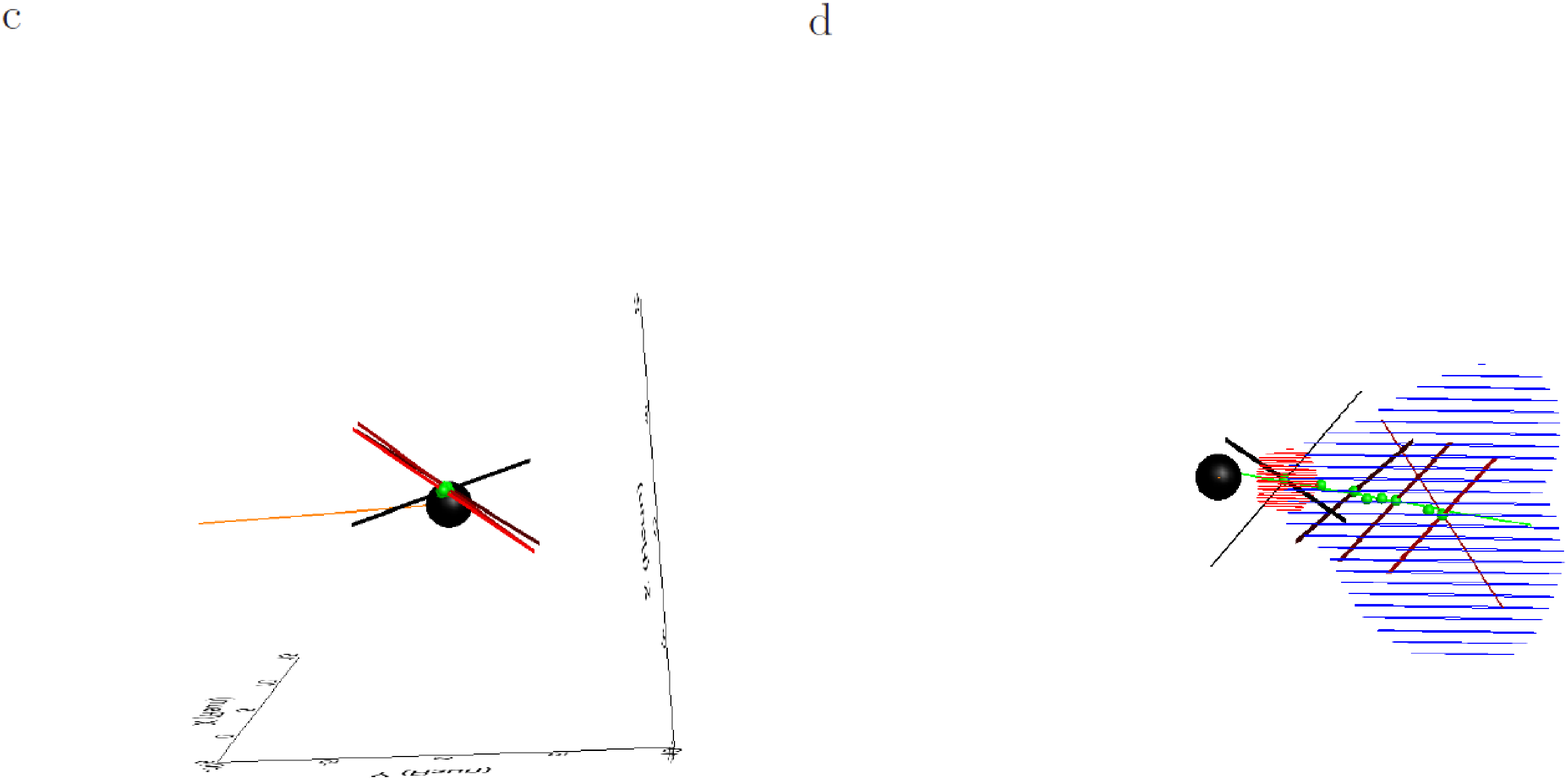}}
  \caption{The orientation of the principle axis at four time instances of 
  18:48, 19:54, 20:39, and 21:24 UT. The color of the principle axis varies from 
  black to red. The green dots are the geometric centre
  of the CME cloud at different times, the green solid line is a 
  linear fit in 3D and indicates the CME propagation direction. The orange solid 
  line lies in the solar equatorial plane and perpendicular to the propagation.
  Panel (a): A view from STEREO A of the intermediate priciple axes together with
  the projected 3D CME at 18:48 and 21:24 UT in red and blue color,
  respectively. Panel (b): A view of the major 
  principle axes from a vantage point above the north pole. Panel (c): A view of 
  the major axes along the propagation direction. Panel (d): A view of the major
  axes along the line in orange and the corresponding projected 3D CME at 18:48 
  and 21:24 UT in red and blue color,respectively.}
  \label{fig:prin_evol}
\end{figure}

By following the variation of the calculated eigenvalues and eigenvectors, we
can deduce the CME's morphological evolution. In Table~\ref{tab:eigen}, the
ratio of eigenvalues of the major to the minor axis, and of the intermediate
to the minor axis are listed. The related uncertainties are propagated from
the 2D CME tracing uncertainties.
In the first column, the average distance between the traced CME peripheries and
its standard deviation are presented. Here the 2D uncertainty is derived by
repeating the tracing process for ten times.
It can be seen that the major morphological differences occurred during the
first three time instances.
After 20:08 UT, the shape appears to have remained stable. This trend can also
been seen from the orientation of principle axes in
Figure~\ref{fig:prin_evol}.

We have to admit that our reconstruction is subject to another uncertainty.
From the coronagraph observations we can only observe the part of CME outside
of the occulter. This produces a larger uncertainty on the reconstructed CME
shape in the early phase of propagation when an unnegligible part of CME was
still below the occulter.
Therefore, the orientation of the reconstructed CME cloud at the earliest
time in Figure~\ref{fig:prin_evol} might have relatively big error bars.

In panel (a) of Figure~\ref{fig:prin_evol}, the intermediate principle axis at
four different time instances is shown from viewpoint of STEREO A. The black
sphere represents the Sun, the red sphere represents STEREO A. The green dots
are the geometric centre at different times. And the green line is the
linear interpolation of these green dots indicating the CME propagation
direction.
The projected CME cloud at 18:48 UT and 21:24 UT is presented as well in red
and blue in panel (a), respectively. We find that the orientation along 
which the CME is most extended as seen in COR A is more or less consistent with 
the intermediate axis, not with the major axis. It implies that for this CME, the
most elongated direction in 3D is not lying in the projected COR A image
plane.

The evolution of the intermediate axis shows big changes in orientation from
18:48 UT to 20:39 UT. This shape change is also evident in COR A observations.
In the early phase of the propagation, the northern part was moving ahead of
the southern part. The reason is that the northern part of the filament
related to this CME erupted first \citep{Reddy:etal:2011}. Later on, the
southern part of the CME entered the fast solar wind region and was probably
accelerated. The interaction of the CME with the fast and slow solar wind
components likely produced a heart-like shape. The slow solar wind 
appears to originate from a region that is known as the streamer belt, which 
was around the solar equatorial plane as can be seen in panel (e) of
Figure~\ref{fig:cme_ini}. Because the CME had a speed (see \S 4.4) greater 
than the slow solar wind, say typically about 400 km~$s^{-1}$, it became
significantly decelerated near the streamer belt; whereas the part of CME 
in higher latitudes might even be accelerated, if the ambient fast solar wind 
had a higher speed. Varying conditions at different latitudes 
likely caused the distortion of the CME structure.
It can also be seen in panel (a) that the
propagation of the CME geometric centre points more or less to the dip of the
concave inward bulge of the CME shape which marks the position of the 
streamer belt.

Panel (b) shows the major axis projected onto the solar equatorial plane.
The thin orange solid line is the direction perpendicular to the CME
propagation. We find that the orientation of the major axis does not show much
variation during the CME evolution. Interestingly, it is not perpendicular to the
propagation direction. The asymmetry with respect to the propagation direction
probably reflects a west-east asymmetry of the filament eruption.

Panel (c) presents the evolution of the major axis as seen along the
propagation direction. Again we find an almost constant orientation during the
later phase of the CME evolution. Our analysis indicates a rotation of the
major axis in the early stage from 18:48 UT to 19:54 UT. However, this rotation
might also be due to the uncertainty of the CME shape at 18:48 UT introduced by
the part of the CME still hidden behind the occulter. A comparison of the
orientation of the CME with the related filament and in-situ ICME will be
discussed in \S 4.3.

The last panel in Figure~\ref{fig:prin_evol} shows the intermediate and the
major axes together with the projected CME cloud as seen from the perspective
perpendicular to the CME propagation at 18:48 UT and 21:24 UT. The major axis 
at 19:54 UT and 20:39 UT from
this perspective are plotted as well. It is clear that the
shape has changed significantly from 18:48 UT to 21:24 UT. Again, the most obvious
change is during the early phase of the evolution.

\subsection{The source region and orientation comparison}

Extrapolating of the CME geometric centre propagation direction backwards to
the solar surface, we arrive at a location very close to AR 11093. Considering
the error bar of the 3D reconstructions especially at the early phase of the
propagation, the active region may be regarded as the possible 
source region. The extrapolation of the geometric centre is quite helpful 
for the identification of a CME's source region, especially
around the solar activity maximum when there are multiple active regions and
more frequent activities.

\begin{figure} 
 \centering
 \vbox{
 \includegraphics[width=10.cm, height=10.cm]{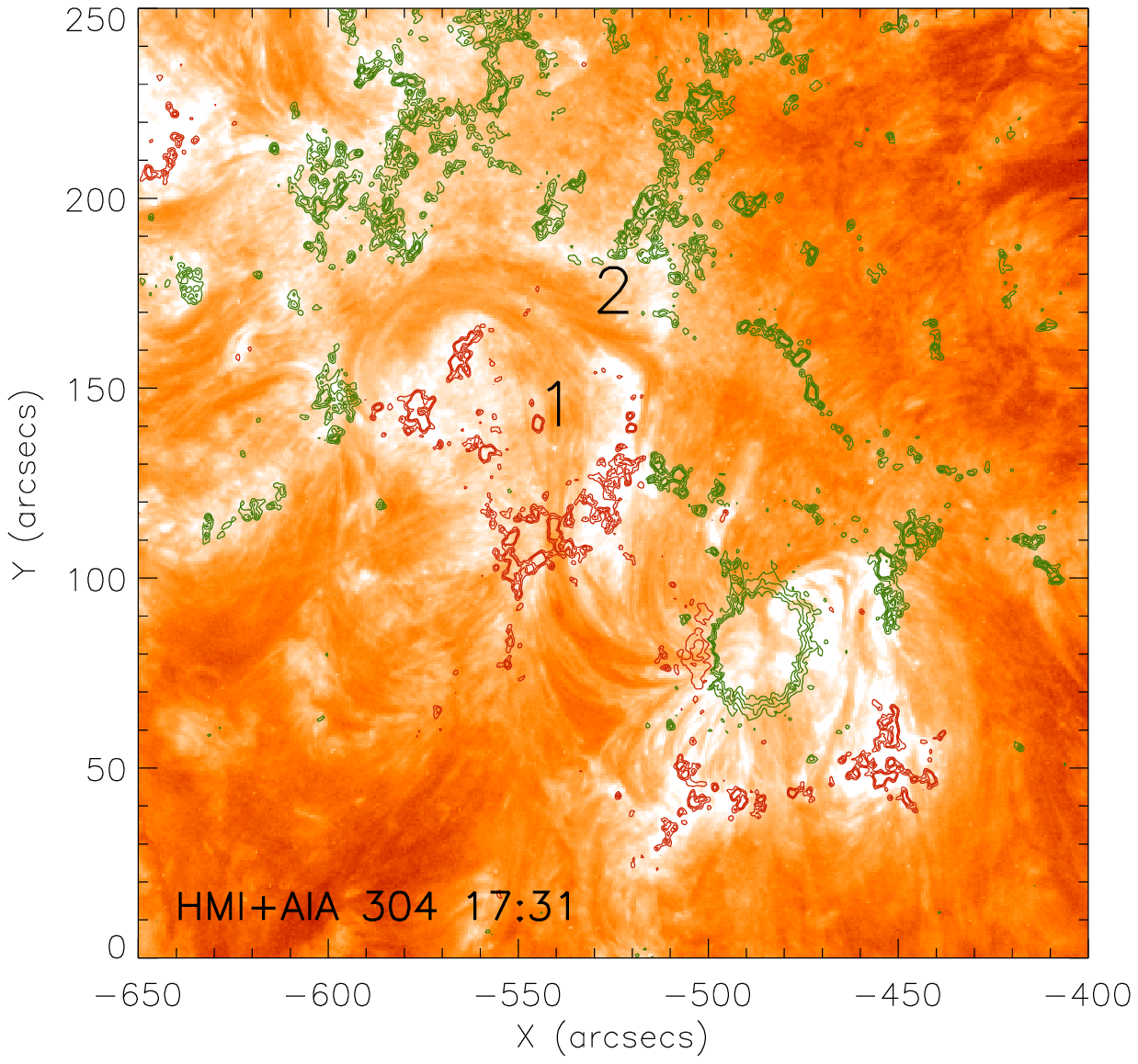}
 \includegraphics[width=10.cm, height=10.cm]{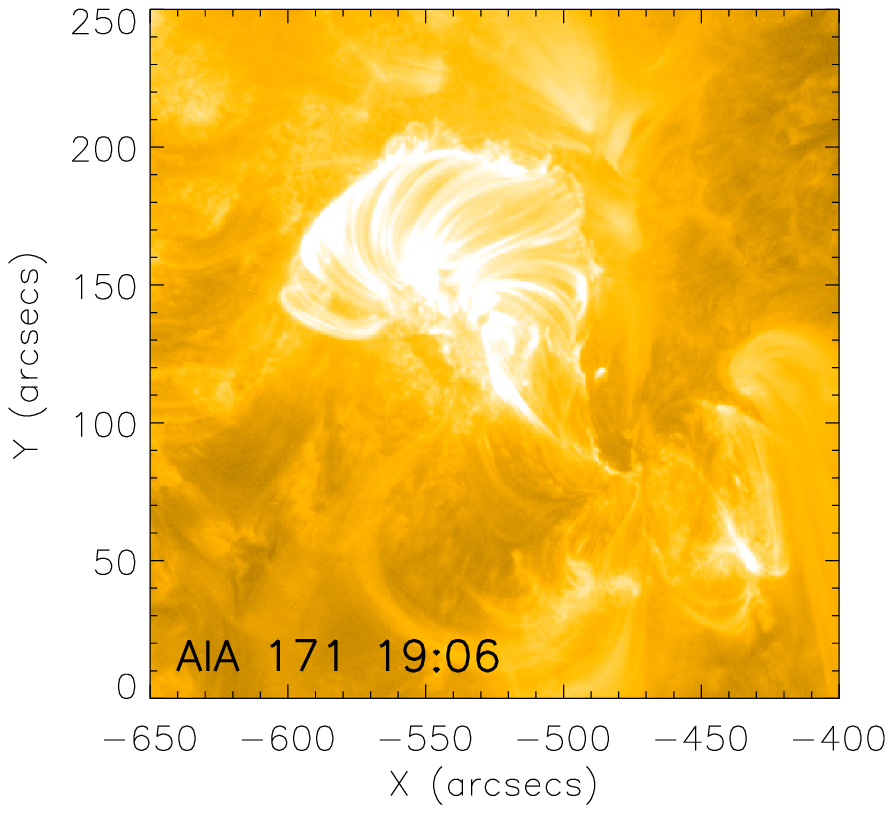}}
 \caption{Upper: AR11093 at 17:31 UT before the flare occurrence. 
  It is recorded by AIA 304~\AA~and HMI magnetogram. The background is the 304 
  image and the red and green contours are the positive and negative polarities 
  in HMI. Bottom: the post flare arcades observed at 171~\AA~by AIA.}
 \label{fig:sourcer}
\end{figure}

Reddy et al. (2011) and Srivastava et al. (2011) made detailed analyses of AR
11093. They found that it was the rising of a filament that led to a M 1.0
class flare and the CME on August 7, 2010. Here we present some
complementary investigations of the source region. In Figure~\ref{fig:sourcer}
are AIA and HMI observations. The upper panel shows the EUV image
at 304~\AA~right before the flare with HMI magnetogram contours overplotted.
The lower panel shows the post-flare arcades observed at 171~\AA~by AIA well
after the CME release.

From GOES observations we find that the flare started around 17:55 UT and peaked
at 18:24 UT. The filament marked 1 in Figure~\ref{fig:sourcer}, upper panel,
rose before the flare which is also indicated by the deviation of this
filament to the magnetic polarity inversion line (PIL). As filament 1 is
relatively high, the projected filament is, due to the projection effect, not
lying along the PIL. However, filament 2 seems to be consistent with the PIL.
Its disappearance is probably related to the heating by the flare.

The orientation of the PIL in the active region approximated by filament 2 can
be compared with the orientation of the major principle axis calculated from
the 3D CME cloud and the flux rope orientation deduced from the VEX/MAG data.
We note that the orientation derived from the VEX data only involves 
the flux rope component in a CME, while the major axis of the reconstructed
3D CME is related to the leading edge, flux rope and possibly some material following
it. In Section 2 we found that the CME flux rope in the coronagraph image 
occupied most of the area which we used for the 3D reconstructions, the major
axis is used as a first approximation of the orientation of the 3D CME
flux rope. 

\begin{figure} 
  \centering
  \includegraphics[width=13.cm, height=10.cm]{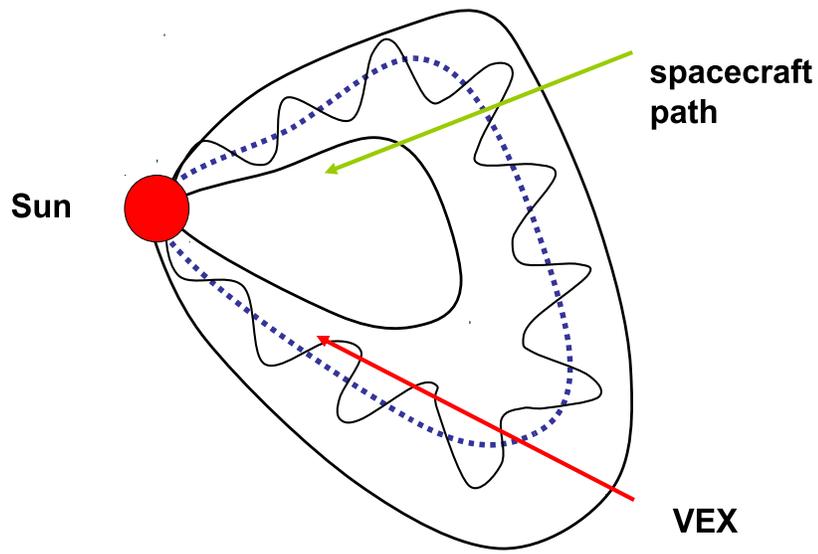}
  \caption{Skematic view of the magnetic flux rope with respect to the spacecraft
  position. The blue dotted line indicates the flux rope axis. The red line 
  represents a possible path of VEX in the flux rope. The green line is an 
  arbitrary path. }
  \label{fig:mc_ske}
\end{figure}

The filament is curved as seen from Figure~\ref{fig:sourcer}, we can
only make a rough comparison. The general orientation of the filament lies in
the direction from northeast to southwest which is consistent with the major
axis indicated in panel (c) of Figure~\ref{fig:prin_evol}. The orientation of
the flux rope observed by VEX was computed by the Minimal Variance Analysis
(MVA) method. It has a direction of (0.96, 0.01, -0.29) in RTN coordinate
system which means that the detected flux rope is mainly in the radial
direction. Compared with the flux rope orientation in the remote-sensing data,
there is a big difference. From the top view of the 
reconstructed 3D CME in Figure~\ref{fig:topview}, it seems realistic that only 
the leg of the flux rope structure passed through VEX. In
consequence, the orientation measured by MAG indicates a nearly radial direction.
If a spacecraft traveled through different parts of the magnetic cloud, the flux rope 
orientation seen by the spacecraft may vary significantly. A diagram depicting 
the possible VEX spacecraft path in the magnetic flux rope is shown in 
Figure~\ref{fig:mc_ske}. However, if the spacecraft went through the flux
rope along the green path, we might arrive at a flux rope axis almost perpendicular
to the green direction.

\subsection{Propagation direction, height, and speed evolution }

\begin{figure} 
  \centering
  \includegraphics[width=9.cm, height=20.cm]{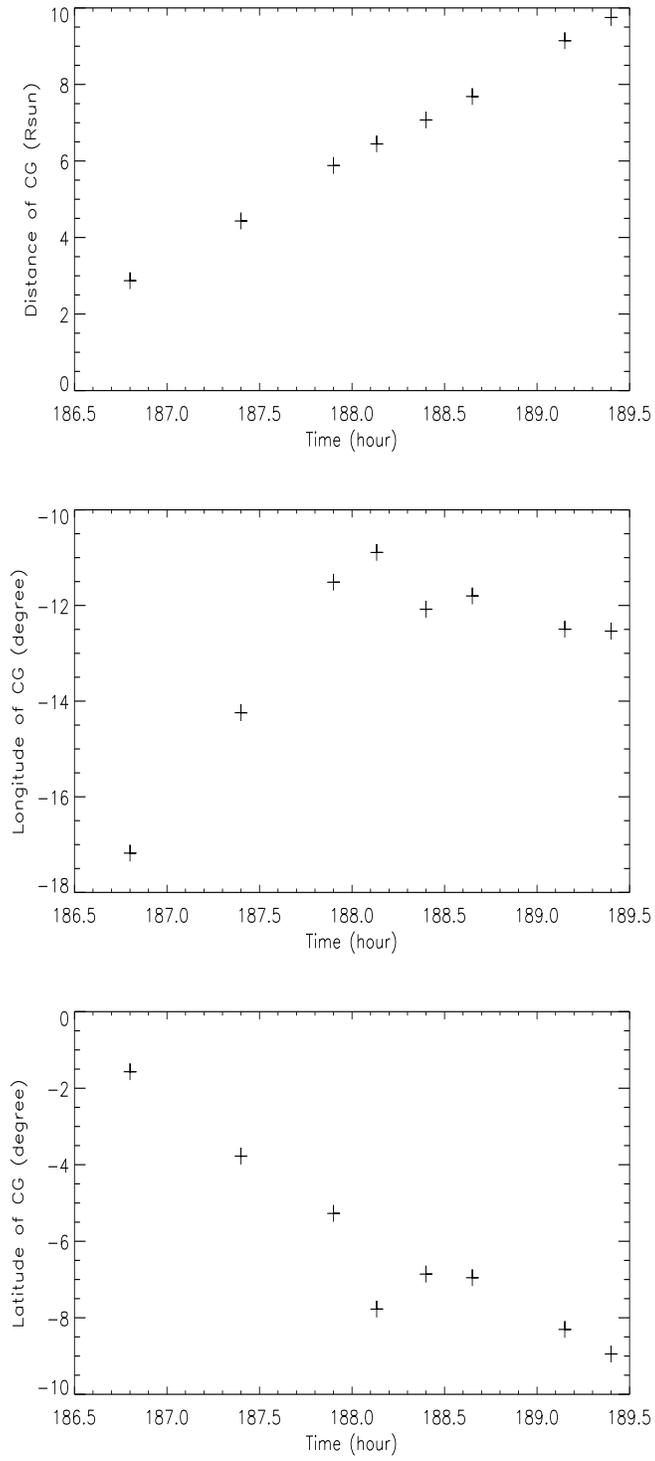}
  \caption{From top to bottom are the time evolution of the distance of
  GC to the Sun centre, the longitude and latitude of GC in units of degree.}
  \label{fig:lonlat_evo}
\end{figure}

In Figure~\ref{fig:lonlat_evo}, the trajectory of the CME 
geometric centre is plotted. The figure shows its distance from the Sun and
its longitude and latitude. Near the Sun, the geometric centre has an
average speed near around 512 $\mathrm{km~s}^{-1}$.
The averaged longitude and latitude in the Carrington coordinates are -12.8
and -6.18 degrees, respectively. The CME was propagating roughly in the radial
direction. However, the latitude decreases from the source region at around 12
degrees in the northern hemisphere to -9 degrees in the southern hemisphere
at a distance of about 10 solar radii.
Similar deflections of a CME have been reported by
\citet{Cremades:Bothmer:2004} and other authors. CMEs which are launched
inside the solar activity belts at a time before the solar maximum and which
possess a total plasma and magnetic pressure less than the ambient solar wind,
are often deflected to the latitude where the slow solar wind resides.

\begin{figure} 
  \centering
  \hbox{
  \includegraphics[width=8.cm, height=7.cm]{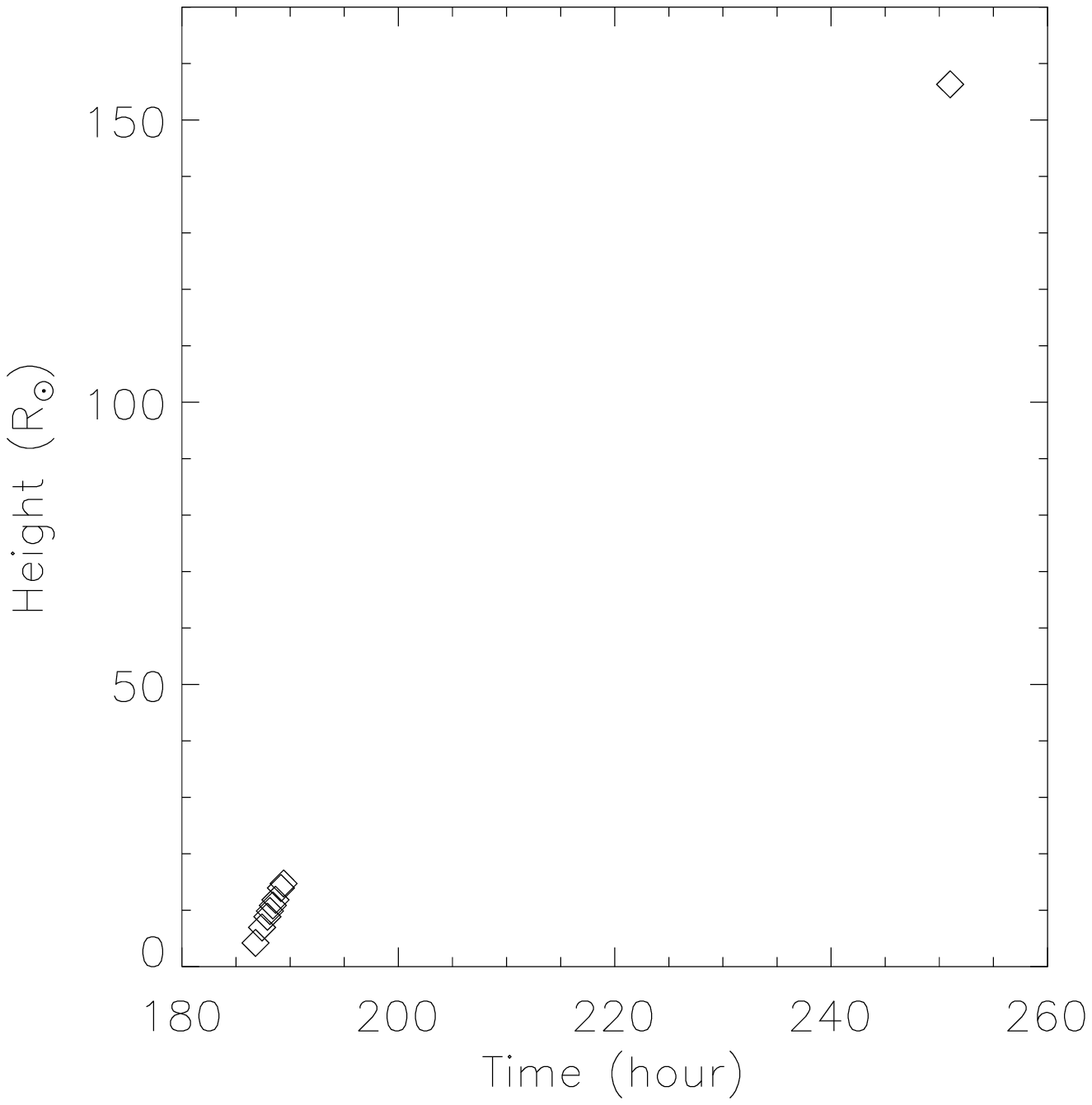}
  \includegraphics[width=8.cm, height=7.cm]{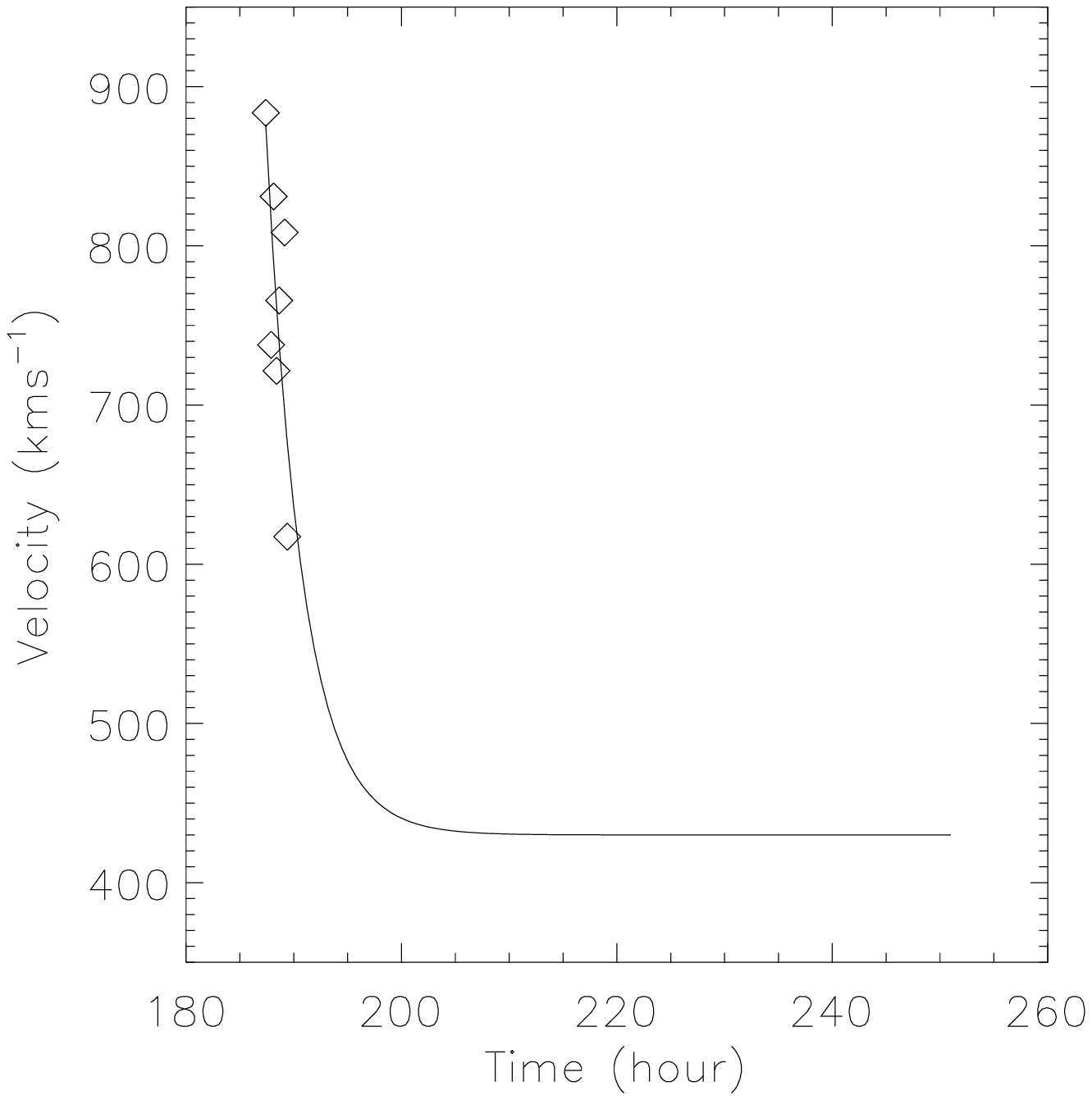}}
  \caption{Left: Height-time profile of the part of CME periphery heading to Venus. 
   Right: the corresponding speed profile with an exponential function fitting 
   to it.}
  \label{fig:ht_vt}
\end{figure}

After deriving the 3D localisation of the CME cloud, we also localise the part
of the CME leading edge around the solar equatorial plane directed to Venus.
In Figure~ \ref{fig:ht_vt}, the corresponding height-time (HT) plot is
presented in the left panel. The last point in the HT diagram is derived from
VEX data. It corresponds to the arrival of the ICME at Venus at around 11:00 UT on
August 10. The
$x$ axis is in units of hours starting from the beginning of August. On the
right side of Figure~\ref{fig:ht_vt}, the speed is calculated from the HT
plot with an exponential fit. The fit is constrained by HT data on the left.
The VEX data point was included such that the respective travel time of the
speed agrees with the distance to Venus.
We conclude from the low asymptotic speed of about 430~$\mathrm{km~s}^{-1}$
that the CME was embedded in the slow solar wind which severely decelerated
the CME.

\section{Conclusions and discussions}


We have developed a new method to obtain the 3D shape of a CME cloud without
assuming a predefined family of shape functions. We applied our method to a
CME which erupted on August 7, 2010. The geometric centre, three principle
axes and the corresponding scale along them were derived and their evolution
with time is presented. We could observe the evolution of the CME shape for
approximately 3 hours. We find that a significant deformation occurred during
the first two hours. During the last hour of observation, the shape evolution
was more limited and continued almost self-similarly. We attribute the change
of the CME morphology during the first two hours to its interaction with the
ambient solar wind. The projected CME major axes are found not
perpendicular to the propagation direction. It is probably due to 
the east-west asymmetry of the CME related filament eruption.

The determination of the CME shape is helpful to interpret the in situ
observations by VEX/MAG, STEREO A \& B/IMPACT, STEREO A \& B/PLASTIC and ACE.
VEX detected a magnetic cloud following the preceding shock, STEREO B only saw
the shock. This is supported by our reconstruction which, if extrapolated out
to 1 AU, predicts that the CME just missed the location of STEREO B. For ACE,
the shock had probably dissolved into the background solar wind before it
arrived at the Earth. Therefore, ACE did not detect any shock signature.

By extrapolating the CME geometric centre propagation direction backwards to
the solar surface, we find that the source region of this CME is likely the
active region AR 11093 located at N12E31 in which a M1.0 class flare occurred
about 40 min before the CME became visible in the STEREO coronagraphs.
The rise of the destabilised filament led to CME eruption. The
orientation of the polarity inversion line in AR 11093 is in general
consistent with the major principle axes of the CME shape we have obtained
from its reconstructed shape.
However, it is not consistent with the flux rope axis deduced from the VEX
data with the MVA method. According to the reconstructed CME localisation in
the equatorial plane, it is very probable that only the leg of the CME flux
rope passed the VEX spacecraft. Indeed, the flux rope observed showed a
strong inclination in the radial direction.

A comparison of the latitude of the source region with the latitude of 
geometric centre of the reconstructed 3D CME implies there 
is a southward component of
its motion. If the plasma pressure and magnetic pressure in the CME is less
than that of the background solar wind, the CME is often deflected away from
the coronal hole where the fast solar wind dominates.

Combining the remote-sensing measurement and the in situ data from VEX, we
derive the 3D distance and speed evolution of the CME motion from Sun to
Venus. The CME speed decreases from 900 $\mathrm{km~s}^{-1}$ at one hour
after its initiation to 430 $\mathrm{km~s}^{-1}$. Probably
the CME speed was finally adjusted to the ambient solar wind flow.

\citet{Temmer:etal:2011} investigated at which heliospheric distance the drag
force starts to prevail over the driving magnetic force so that the speed of
the ICME gets finally adjusted to the speed of the ambient solar wind. They
analysed the observations of different CMEs and found that this heliospheric
distance varies considerably from below 30 $R_\odot$ to beyond 1 AU. 
Theoretical and observational work on the driving and drag
forces during the CME evolution and propagation from the Sun to the 
interplanetary medium can also be found in \citet{Chen:1996}, 
\citet{Manoharan:etal:2001}, \citet{Tappin:2006} and \citet{Howard:etal:2007}. 
For the
CME we have analysed, we estimate that this distance at which the CME speed
becomes adjusted to the solar wind speed is beyond the FOV of COR2. Therefore,
to derive a reliable speed profile, eventually to make an accurate
prediction of the CME transit time on Earth or other planets, at least
Heliospheric Imager I (HI 1) in the SECCHI instrument package has to be
included for the investigated CME. In the near future, we will extend the mask
fitting method to HI observations.

A precise prediction of the CME travel time also depends on the accuracy of
the reconstruction method. Comparison of our mask fitting method and other 3D
reconstruction methods, for example, GCS forward modeling, polarisation ratio,
local correlation track combined with tie-pointing, geometrical localization,etc., 
will be presented in another paper. For this particular CME associated
with a complicated polarity inversion line, the forward modeling of the CME by
a single flux rope model will probably be not very successful owing to the odd
shape of the CME. A fit with two flux ropes might give better results. A
shortcoming of our mask fitting method is that the internal structure of the
CME is not included. We are planning to extend our shape reconstruction method
by a tomographic approach to determine a constrained density distribution
model within the 3D CME cloud. 

\acknowledgements

We thank CDAweb for making the STEREO/IMPACT, STEREO/PLASTIC data available. 
Thanks Marilena Mierla for many discussions on this work during her visit in the Max-
Planck-Institut for Solar System Research.
STEREO is a project of NASA. The SECCHI data used here were produced by an
international consortium of the Naval Research Laboratory (USA),
Lockheed Martin Solar and Astrophysics Lab (USA), NASA Goddard Space
Flight Center (USA), Rutherford Appleton Laboratory (UK), University of
Birmingham (UK), Max-Planck-Institut for Solar System Research (Germany),
Centre Spatiale de Li\`ege (Belgium), Institut d'Optique Th\'eorique et
Applique\'e (France), Institut d'Astrophysique Spatiale (France).
LF and WQG are supported by grants 10833607, 11003047 and 11078025 
from NSFC and 2011CB811402 from MSTC.
The contribution of BI benefitted from support of the German Space
Agency DLR and the German ministry of economy and technology
under contract 50 OC 0904.


\begin{table} 
\begin{center}
 \begin{tabular}{cccccc}
\tablecolumns{6}
Time & 2D uncertainty(arcsec) & major/minor & intermediate/minor\\
\hline\hline
18:48 & $85.8\pm22.15$  & $2.43\pm0.111$ & $1.68\pm0.066$\\
19:24 & $133.9\pm16.91$ & $2.02\pm0.087$ & $1.71\pm0.096$\\
19:54 & $115.1\pm18.34$ & $2.35\pm0.104$ & $1.70\pm0.095$\\
20:08 & $116.7\pm16.54$ & $2.11\pm0.075$ & $1.53\pm0.057$\\
20:24 & $112.5\pm17.47$ & $2.13\pm0.073$ & $1.52\pm0.048$\\
20:39 & $125.0\pm25.05$ & $2.01\pm0.086$ & $1.52\pm0.071$\\
21:08 & $155.8\pm21.27$ & $1.90\pm0.109$ & $1.50\pm0.052$\\
21:24 & $143.3\pm16.12$ & $1.95\pm0.054$ & $1.56\pm0.075$ \\
\hline
\end{tabular}    
\end{center}
\caption{Eigenvalue analysis at different time instances. The first column
shows the average distance between the traced 2D CME peripheries and its standard 
deviation. The second column is the ratio of eigenvalues between the major 
and minor axes, and the third column is the ratio of eigenvalues between
the intermediate and minor axes. The values after the positive-minus sign
are the standard deviation of the ratios. The uncertainty is propagated from 
the uncertainty of 2D coordinates.}
\label{tab:eigen}
\end{table}

\end{document}